\documentclass[conference]{IEEEtran}
\IEEEoverridecommandlockouts
\usepackage{cite}
\usepackage{booktabs}
\usepackage{url}
\usepackage{multirow}
\usepackage{enumerate}
\usepackage{mathtools}
\DeclarePairedDelimiter\ceil{\lceil}{\rceil}

\usepackage{cite}
\usepackage{amsmath,amssymb,amsfonts}
\usepackage{graphicx}
\def\BibTeX{{\rm B\kern-.05em{\sc i\kern-.025em b}\kern-.08em
    T\kern-.1667em\lower.7ex\hbox{E}\kern-.125emX}}
    
\begin{document}
%
\title{Blockchain Enabled Secure Authentication for Unmanned Aircraft Systems}


\author{%
    Yongxin Liu$^{1*}$, Jian Wang$^{2*}$, Yingjie Chen$^{3}$, Shuteng Niu$^{4}$, Zhihan Lv$^{3}$, Lei Wu$^1$, Dahai Liu$^{2}$, Houbing Song$^{2}$\\
    $^{1}$Auburn University at Montgomery, AL 36117 USA\\
    $^{2}$Embry-Riddle Aeronautical University, Daytona Beach, FL 32114 USA\\
    $^{3}$Qingdao University, Qingdao, Shandong 266071 China\\
    $^{4}$Bowling Green State University, Bowling Green, OH 43403 USA\\
    $^{1}$\{yliu19,lwu\}@aum.edu, $^{2}$wangj1@my.erau.edu, dahai.liu@erau.edu, h.song@ieee.org\\
    $^{3}$2018207163@qdu.edu.cn, lvzhihan@gmail.com, $^{4}$sniu@bgsu.edu
}

\markboth{IEEE Internet of Things Journal,~Vol.~11, No.~4, May~2021}%
{Shell \MakeLowercase{\textit{et al.}}: Bare Demo of IEEEtran.cls for Journals}

\IEEEtitleabstractindextext{%
\begin{abstract}
The integration of air and ground smart vehicles is becoming a new paradigm of future transportation. A decent number of smart unmanned vehicles or UAS will be sharing the national airspace for various purposes, such as express delivery, surveillance, etc. However, the proliferation of UAS also brings challenges considering the safe integration of them into the current Air Traffic Management (ATM) systems. Especially when the current Automatic Dependent Surveillance Broadcasting (ADS-B) systems do not have message authentication mechanisms, it can not distinguish whether an authorized UAS is using the corresponding airspace. In this paper, we aim to address these practical challenges in two folds. We first use blockchain to provide a secure authentication platform for flight plan approval and sharing between the existing ATM facilities. We then use the fountain code to encode the authentication payloads and adapt them into the de facto communication protocol of ATM. This maintains backward compatibility and ensures the verification success rate under the noisy broadcasting channel. We simulate the realistic wireless communication scenarios and theoretically prove that our proposed authentication framework is with low latency and highly compatible with existing ATM communication protocols.

\end{abstract}

}

\maketitle

\IEEEdisplaynontitleabstractindextext

%
\IEEEpeerreviewmaketitle

\section{Introduction}
%
%
%
%

The emerging Unmanned Aircraft System stimulates the development of a new paradigm of transportation, the integration of different modes of autonomous and smart vehicles to facilitate the agile mobility of people and cargo between places \cite{song2016cyber, song2017smart, song2021security}. In this context, Advanced Air Mobility (AAM) aims to develop air transportation systems that move people and cargo between places previously not served or underserved by aviation using revolutionary new aircraft, such as smart UAS \cite{hackenberg2019nasa,HASSIJA202051}. The principal requirement for AAM is the coexist of UAS and regular aircraft within the national airspace under a unified and coordinated management framework \cite{8337902, 9171476, song_liu_wang_2021}. However, this requirement brings challenges to the current Air Traffic Management (ATM) systems. Currently, several obstacles impede the integration of UAS into the national airspace, and the most difficult ones are real-time verification and airspace usage authentication \cite{7452265, TIAN2019102354,9440979}. These criteria require the UAS to broadcast its identity to ATM facilities in a verifiable and trustworthy way. 

In regular scenarios, each operational aircraft uses an ADS-B transponder at 1090MHz to broadcast its ID, and other critical flight information to the ATC facilities \cite{riddle1090,9488323} while a Remote ID Broadcast Module \cite{UASRemot21:online,9237145} is required for UAS. In these solutions, all communication entities within the current solution broadcast the critical information using plain text without cryptographic features. Consequently, ATM systems are still vulnerable to cyber attacks. For example, a malicious adversary can deliberately broadcast a high volume of fake data packets to mislead and interrupt the normal operations of ATC. In addition, the ADS-B protocol has become a de facto standard with corresponding transponders installed on nearly every aircraft. It is impossible to upgrade them in a short time for secure authentication features. System upgrades in the aviation industry need to be taken with sufficient backward compatibility.

Non-cryptographic solutions have been proposed to facilitate the existing ATM systems with identity verification capability\cite{liu2020zero,liu2020deep,9425491}. The key idea of these approaches is to derive the protocol-agnostic physical layer features from the raw signals for pattern recognition, and transmitter identification \cite{8531759,9492269,9377975}. However, on the one hand, the physical layer features of raw signals can change drastically under different conditions (e.g., interference, weather, or antenna characteristics), making it difficult to get a widely acceptable wireless transmitter identification model. On the other hand, although cryptographic identity verification approaches are more reliable, the communication channel bandwidth between UAS and ATM facilities is limited. Therefore, special mechanisms are needed to ensure the successful delivery of authentication information while maintaining backward compatibility with the existing ATM infrastructures.

In this paper, we propose a secure framework for the safe integration of UAS into the national airspace. We first propose a blockchain enabled framework for the authentication of flight plans and exchanging secret keys between UAS and ATM infrastructures. We then provide an enhanced ADS-B protocol using fountain code \cite{8711001,mackay2005fountain} for the reliable transmission of authentication information in capacity-limited wireless channels. Our contributions are as follows:
\begin{itemize}
    \item We provide a blockchain based framework for the review and authentication of flight plans between UAS and existing ATM infrastructure, which is reliable and robust to single-point failure under cyber attacks.
    \item We provide a backward compatible and enhanced ADS-B protocol for aviation communication. The protocol increases the trustworthiness and reliability of communication between UAS and ATM.
    \item We theoretically prove that our proposed authentication framework is highly compatible with the de facto ATM communication protocol.
\end{itemize}
Our research offers a solution for cyber defense in air and ground smart vehicle networks. This solution maintains the best backward compatibility with the existing air traffic management facilities and can be extended to protect the communication channel of manned aircraft. The remainder of this paper is organized as follows: A literature review of related works is presented in Section~\ref{sectRW}. 
The methodology in Section~\ref{sectMM}. Performance evaluation is presented in Section~\ref{sectEED} with conclusions in Section~\ref{sectCC}.
\addtolength{\topmargin}{0.01in}

\section{Related Work}
\label{sectRW}
\subsection{Blockchain in Cyber Physical Systems}
Blockchain is a specific kind of decentralized and distributed database that utilizes cryptography algorithms and consensus mechanisms to prevent information from being modified maliciously \cite{CHEN20191122, 9199797}. The most elegant feature of the blockchain is information integrity, in which no one can modify the data without controlling greater than 50\% of computational resources among all peers \cite{8048633,ALLADI2020100249}. Blockchain has been adapted to many Cyber-Physical Systems for secure data storage \cite{khan2018iot}. For example, in \cite{novo2018blockchain}, a scalable framework is proposed to manage the accessibility of computational resources using Blockchain. Their approach introduces a Private Blockchain Network in the Cloud as well as one Management Hub. In \cite{sharma2017software}, a Blockchain-enabled hierarchical Cloud architecture for CPS is proposed, and the Blockchain is deployed as a secure information exchange media within the service provider and consumers. Blockchain can also be used as a secure middleware for network routing and authentication \cite{WANG2021131,liu2019blockchain,WANG2021233}.

\subsection{Non-cryptographic Identity Verification in Air Traffic Surveillance Systems}

ATM systems are becoming a crucial infrastructure for VANETs and UAS. In this domain, the non-cryptographic identity verification of UAS and manned aircraft is attracting attentions. Authors in \cite{leonardi2017air} used the instantaneous phase varying pattern of signals as a distinctive feature for aircraft identification. Phase patterns can be applied to low-cost equipment, as in \cite{ying2019detecting} and \cite{chen2019deep}, the authors extract the phase patterns in each ADS-B message and use a Deep Neural Network (DNN) model to identify the genius ones. However, a common problem of these research is that the identification model is not robust to the varying wireless channels and RF front-ends. This motivates us to provide a more robust solution using proactive fingerprint embeddeding with backward compatibility.


\section{Methodology}
\label{sectMM}
This section will provide an overview of the system model, followed by the blockchain infrastructure and procedures for secure authentication of UAS with the existing ATM facilities. We will then introduce our modification of the enhanced ADS-B protocol for the identity verification of UAS in real-time.
\subsection{System Model}
\begin{figure}[h]
    \centering
    \includegraphics[width=1.00\linewidth]{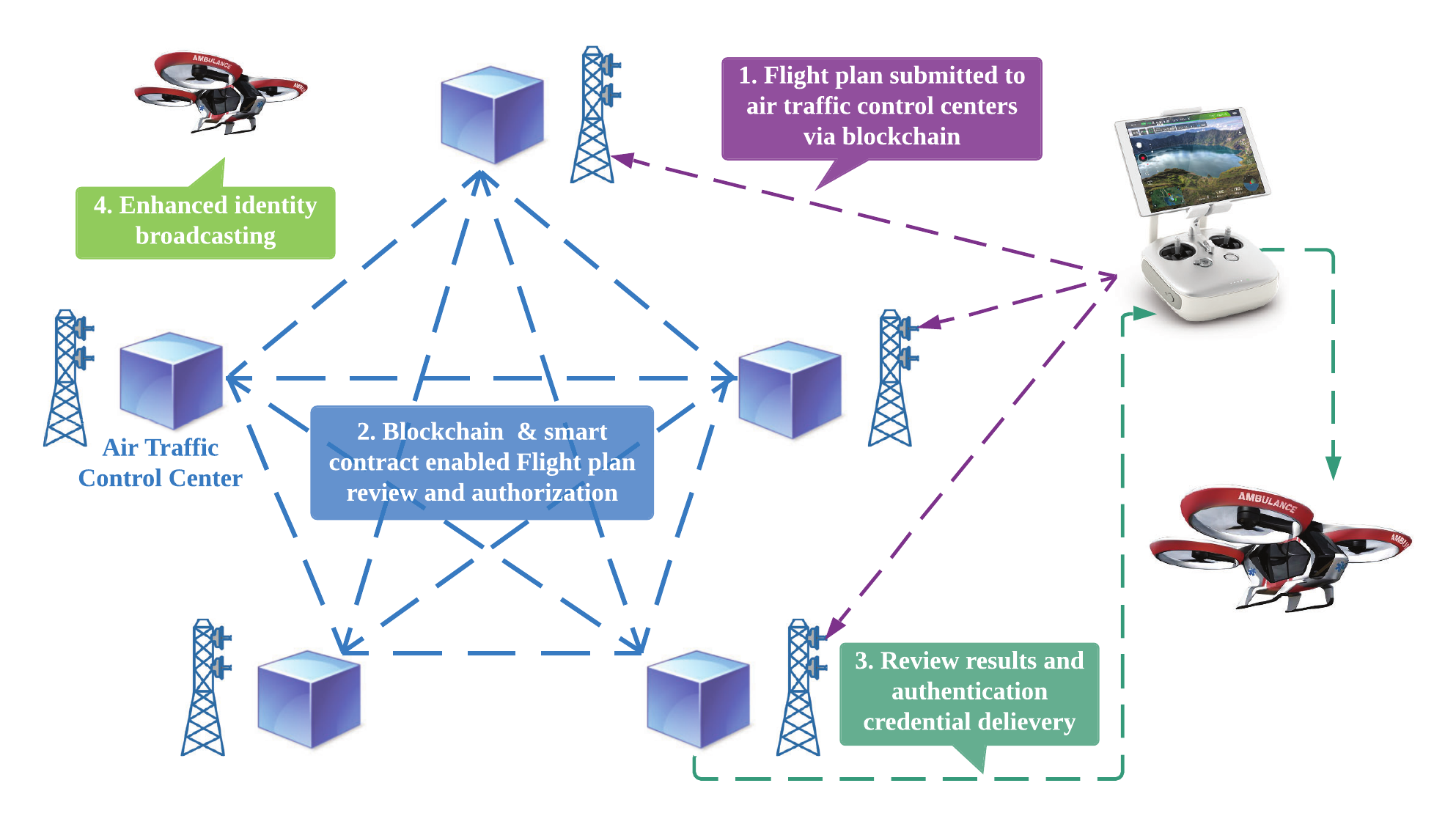}
    \caption{Blockchain authentication framework for UAS}
    \label{figChainATC}
\end{figure}

The system diagram of our proposed solution is in Figure~\ref{figChainATC}. As depicted, we suppose that all ATM facilities (including ATC facilities and weather observation stations) are connected as nodes in a private blockchain network. In this ATM blockchain network, vital information such as airspace availability, weather, flight plans, etc., can be shared and stored securely. We also define that UAS users are supposed to submit their flight plans via regular civilian communication media, such as cellular networks. The submitted flight plans are assessed and authorized by smart contracts. Once a flight plan is approved, the en-route authentication credentials (tokens) will be delivered to the UAS and are utilized to generate the enhanced ADS-B packets.

\subsection{Blockchain for Secure Flight Plan Assessment}
We utilize blockchain and smart contracts to make flight plan assessment transparent and secure. In other words, the malicious needs to compromise at least 51\% of the nodes in the blockchain network to malfunction the assessment process. The procedures of flight plan submission and assessment procedure of flight plans are as follows:
\begin{enumerate}[\textbf{Step} 1:]
    \item UAS users submit flight plans with digital signatures to the ATC blockchain network via ground control stations. Flight plans will be broadcast to all nodes in the ATM blockchain network. In general, a flight plan contains waypoints, estimated times of arrival, and the UAS registration number.
    \item Once a flight plan is broadcast, the smart contract will assess it with the newest regulations and information. If the smart contract approves a flight plan, an appropriate node in the ATM blockchain will be selected to generate an authentication credential for this flight plan. We first generate a list of the top ten candidates using the metric as:
    \begin{align}
        s =\dfrac{1}{D + L + 1}
    \end{align}
    where $L$ is the averaged latency of authentication credentials generation within the past 24 hrs, and $D$ is the distance between the take-off point and the related nodes within the air space. We then randomly select one out of ten to generate the authentication credential. 
    \item The authentication credential of a flight plan is composed of: i) a temporal symmetric key ($SK_p$). ii) a hash digest for the flight plan ($ID_{fp}$) generated by the smart contract. iii) a challenge string. iv) the digital certificate of the flight plan signed by the issuance facility ($ID_{if}$). iv) The LT Code message generation matrix $\boldsymbol{M_{LT}}$ \cite{1181950}:
    \begin{align}
    \label{eqEncMatrix}
        \boldsymbol{M_{LT}}=\begin{bmatrix}
            1 & 0 & 0 &\cdots&\\
            \vdots  & \vdots & \vdots & \cdots &\\
            0 & 1 & 1 & \cdots&\\
        \end{bmatrix}_{K\times N}
    \end{align}
    where $\boldsymbol{M_{LT}}$ is the message generation matrix of LT code, a practical implementation of fountain code. The message generation matrix contains only zeros and ones, $\boldsymbol{M_{LT}}$ needs to have full rank. It actually represents the Tanner Graph of the LT code encoder. LT code is the practical implementation of fountain code.
    \item The authentication credential of a flight plan is first broadcast within the ATM blockchain network. Then it's encrypted by the public key ($PubK_{U}$) of the UAS user and sent back to the ground control station of UAS.
\end{enumerate}
Given these steps, a UAS user can verify that a trustworthy entity formally approves the flight plan. Meanwhile, a UAS will be able to generate its authentication information during its flight. In this work, we assume that the UAS can always maintain a reliable Internet connection in the preflight stage. We believe that temporal symmetric encryption is more appropriate to secure the en-route communication between UAS and ATC facilities with low overhead.
\begin{figure}[h]
    \centering
    \includegraphics[width=1.00\linewidth]{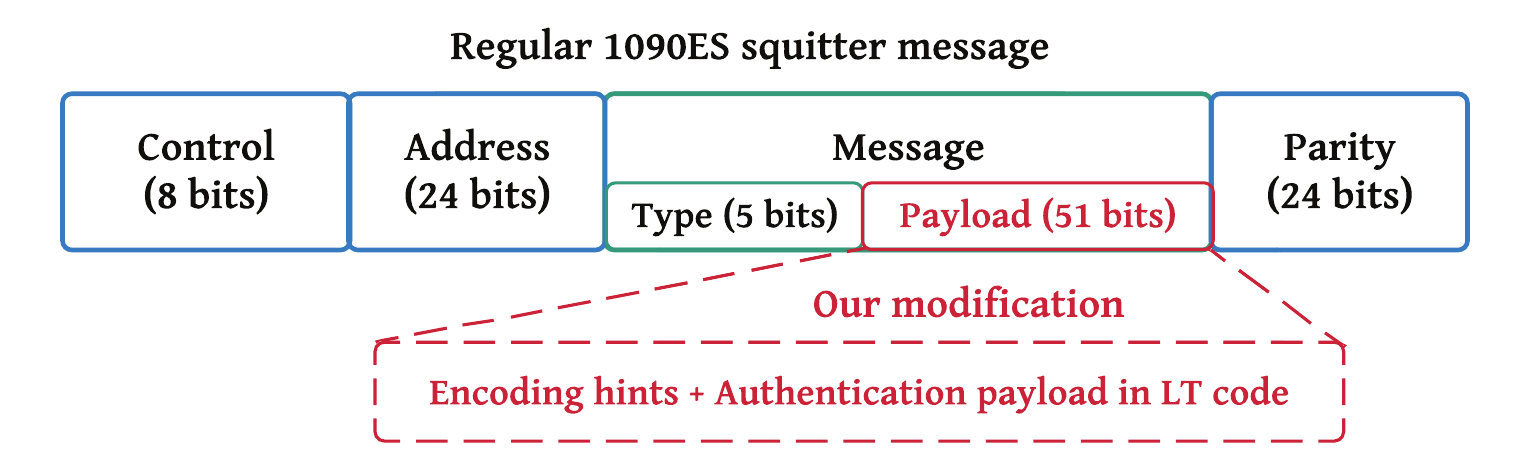}
    \caption{Data block of ADS-B message and our modification}
    \label{figADSBPacket}
\end{figure}
\subsection{Enhanced ADS-B Protocol for Reliable Broadcast Authentication}
\addtolength{\topmargin}{0.01in}

ADS-B is the standard for ATC and ATM. In a Mode S transponder with Extended Squitter, a standard data packet contains only 112 bits and does not contain any authentication payload \cite{riddle1090}. Therefore, some modifications are needed. Specifically, we propose a new message category termed as authentication message. According to the data blocks of ADS-B message in Figure~\ref{figADSBPacket}, we can have 51 bits available in the periodically broadcast 1090ES squitter message. Therefore, the size of the encoded symbols for LT code needs to be less than 51 bits because we also need to reserve bits to indicate how we generate the symbol. These hints are termed as \textit{encoding hints} in Figure~\ref{figADSBPacket}. Given that the length of the raw authentication payload is $M$, we need to divide the raw authentication message into $K$ equal-sized symbols ($C_1,\cdots,C_K$), and the size of each symbol ($L_C$) has to satisfy:
\begin{align}
\label{eqSymbolSize}
    L_C < 51- L_h
\end{align}
where $L_h$ is the size of the encoding hints. In our solution, an encoding hint indicates the column vector in $\boldsymbol{M_{LT}}$ that is used to generate an encoded symbol. We also define that the number of column in $\boldsymbol{M_{LT}}$ needs to satisfy:
\begin{align}
    N > r_0 \cdot \dfrac{M}{L_C}
\end{align}
where $r_0 (r_0 > 1)$ is the redundancy coefficient and $M$ is the size of the authentication payload before encoding. It means that during the transmission of a complete authentication payload, every symbol is generated using a different sub graph. 

To avoid reusing columns in the message generation matrix, we have the following constraint:
\begin{align}
    51 - \ceil*{Log_2\big[N\big]} \geq L_C
\end{align}
where $\ceil*{Log_2\big[N\big]}$ is the length of the encoding hints, $M$ is the size of authentication payload, $L_C$ is the size of encoded symbols. Therefore, the configuration of parameters can be categorized into a constrained optimization problem. 
\begin{figure}[h]
    \centering
    \includegraphics[width=1.00\linewidth]{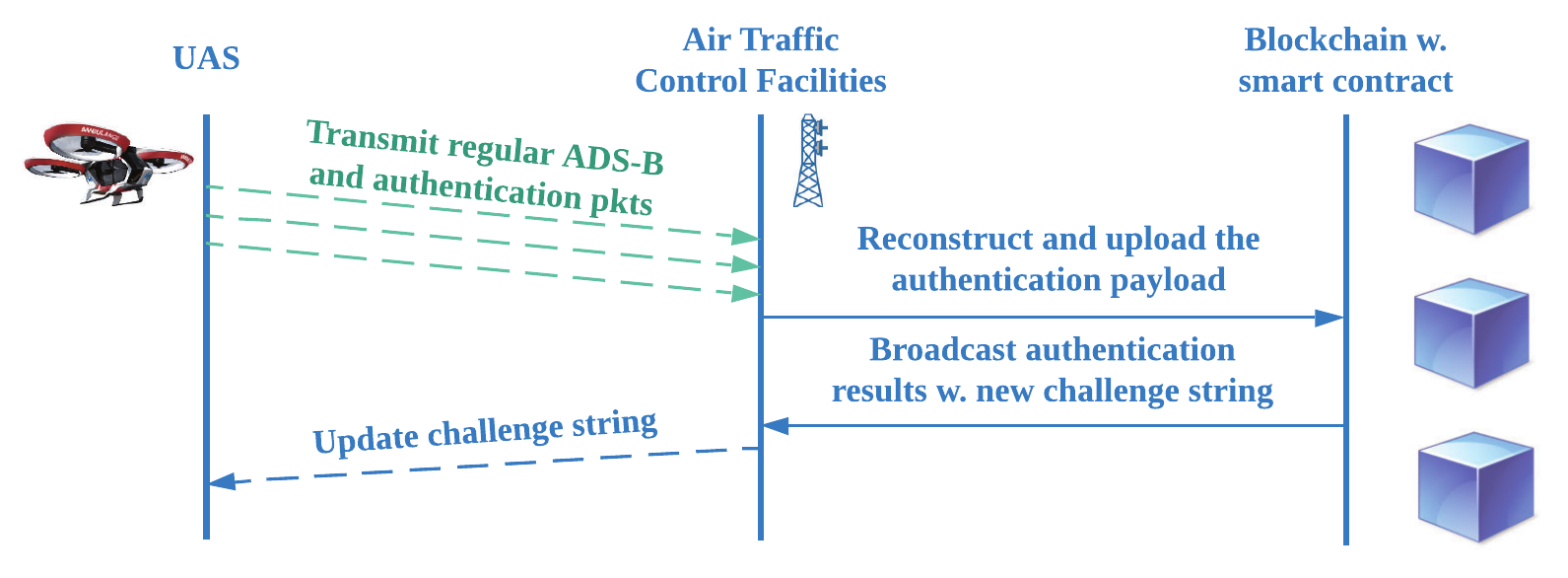}
    \caption{En-route authentication}
    \label{figEnrouteAuthentication}
\end{figure}

Once a UAS takes off, it is supposed to broadcast the authentication payload using LT Code periodically. The authentication payload contains the certificate of the current flight plan ($ID_{fp}$) plus a response to the cryptographic challenge. The authentication procedure is depicted in Figure~\ref{figEnrouteAuthentication}, also as follows:

\begin{enumerate}[\textbf{Step} 1:]
    \item If the LT Code message generation matrix is with $K$ rows and $N$ columns as in (\ref{eqEncMatrix}). The message generation procedure resembles matrix multiplication, but we use binary XOR to replace the binary addition during calculation:
    \begin{align}
        [E_1,\cdots,E_N ] &= [C_1,\cdots ,C_K]\times \boldsymbol{M_{LT}}
    \end{align}
    where $C_1,\cdots,C_K$ are segments of the raw message (the authentication payload). $E_1,\cdots,E_N$ are encoded authentication messages to broadcast. In the context of fountain code, $E_1,\cdots,E_N$ are termed as droplets.
    \item Once a ATC facility intercepts $N_L (M/L_C)$ authentication messages, it will then try to decode and reconstruct the authentication payload:
    \begin{align}
        [C_1,\cdots ,C_K] = [E_1,\cdots,E_N ] \times \boldsymbol{M^{-1}_{LT}}
    \end{align}
    where $\boldsymbol{M^{-1}_{LT}}$ is the inverse of $\boldsymbol{M_{LT}}$. We can also use Belief Propagation algorithm to reconstruct the payload \cite{1181950}. The reconstructed authentication payload will be broadcast and sent to the smart contract.  
    \item The smart contract can broadcast the authentication result and periodically change the challenge string via ADS-B interrogation protocol \cite{ADSBOper24:online}.
\end{enumerate}

For UAS with higher computational capabilities, we can further require the UAS to digitally sign its response to the challenge string. In which it will be computationally impossible to generate fake authentication credentials.

\section{Evaluation and Discussion}
\label{sectEED}
In this section, we will discuss the attack surface of the framework and evaluate the performance of the en-route authentication payload encoding scheme.
\begin{table}[h]
\centering
\caption{The possible malicious attacks and success criteria in our framework}
\label{tabAttackSurface}
\resizebox{\linewidth}{!}{%
\begin{tabular}{@{}cll@{}}
\toprule
\multicolumn{1}{l}{Stage} & Malicious attacks &Attack success criteria \\ \midrule
\multirow{4}{*}{Preflight} & Submiting falsified flight plans. & \begin{tabular}[c]{@{}l@{}}The malicious gains the private key\\ of UAS.\end{tabular} \\ \cmidrule(l){2-3} 
 & \begin{tabular}[c]{@{}l@{}}Malfunctioning flight plan \\ assesssment functionality.\end{tabular} & \begin{tabular}[c]{@{}l@{}}The malicious controls at least one\\ third of the nodes in the ATM\\ blockchain.\end{tabular} \\ \cmidrule(l){2-3} 
 & \begin{tabular}[c]{@{}l@{}}Generating fake flight approval \\and authentication credentials.\end{tabular} & \begin{tabular}[c]{@{}l@{}}The malicious gets the private keys of\\ ATM facilities.\end{tabular} \\ \cmidrule(l){2-3} 
 & \begin{tabular}[c]{@{}l@{}}Intercepting flight plans approvals \\ and authentication credentials.\end{tabular} & \begin{tabular}[c]{@{}l@{}}The malicious gets the private key of\\ UAS or gain the assessibility of ATM \\ blockchain.\end{tabular} \\ \midrule
\multirow{2}{*}{En-Route} & \begin{tabular}[c]{@{}l@{}}Transmitting fake authentication \\ messages.\end{tabular} & \begin{tabular}[c]{@{}l@{}}The malicious gets the authentication\\ credentials or will be impossible if the\\ challenge-response requires digital\\ signature.\end{tabular} \\ \cmidrule(l){2-3} 
 & \begin{tabular}[c]{@{}l@{}}Bypassing the en-route \\authentication mechanism\end{tabular} & \begin{tabular}[c]{@{}l@{}}The malicious needs to control at \\least one third of the nodes in the \\ATM blockchain.\end{tabular} \\ \cmidrule(l){2-3}
 
& \begin{tabular}[c]{@{}l@{}}Recording and re-transmit data\\ packets to create ghost aircraft.\end{tabular} & \begin{tabular}[c]{@{}l@{}}Can temporally succeed but will be \\detected once the challenge string \\is updated and contains a sequence \\number.\end{tabular} \\ \cmidrule(l){2-3}

& \begin{tabular}[c]{@{}l@{}}Compromising a few nodes to inject \\fake en-route information to the \\ATM blockchain.\end{tabular} & \begin{tabular}[c]{@{}l@{}}Depends on how many nodes an \\attacker controls. Since ADS-B \\messages can be picked up by\\ multiple recipients.\end{tabular} \\ \bottomrule
\end{tabular}%
}
\end{table}
\subsection{Security Analysis}

We discuss the possible attacks in two stages: preflight and en-route, as in Table~\ref{tabAttackSurface}. The malicious attackers need to get access to the ATM blockchain or gain the private keys of UAS and its owners to compromise the security mechanism. There are many ways we can prevent malicious attackers from accessing the ATM blockchain, such as multi-factor authentication. Therefore, we believe this authentication framework is sufficiently secure.

\subsection{Performance Evaluation of the Encoding scheme}
\begin{figure}[h]
    \centering
    \includegraphics[width=0.8\linewidth]{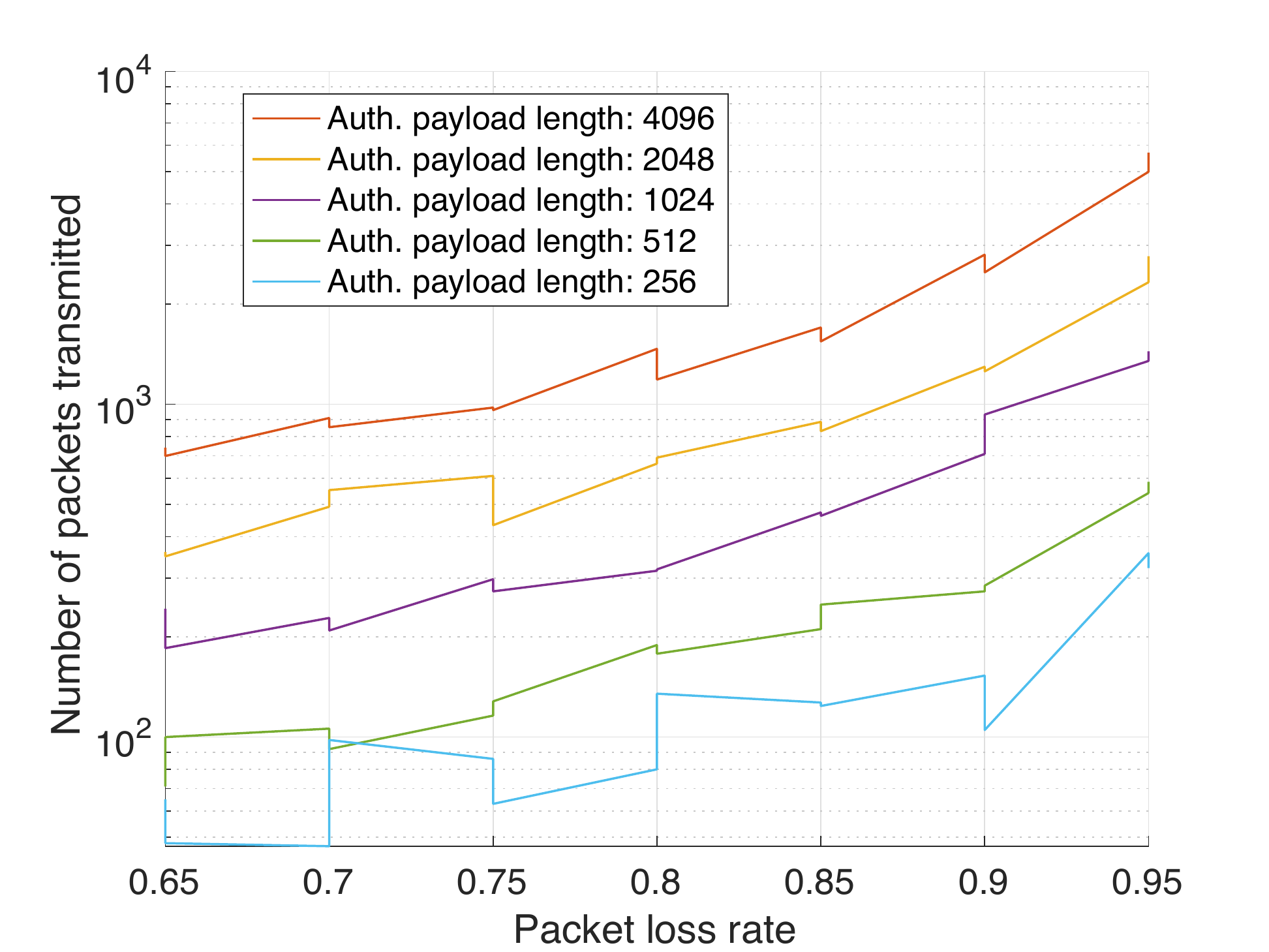}
    \caption{Average number of packets needed to reconstruct authentication payloads with symbol size being 16 bits.}
    \label{figPLRvsNeededPkts}
\end{figure}

For ADS-B systems, the worst-case Packet Loss Rate (PLR) with distance has been proposed in \cite{sciancalepore2019reliability} as:
\begin{align}
    y = 65.366 + 0.020319\times distance[km] 
\end{align}
Therefore, we only need to set the range of PLR from 65\% (less than 1 km) to 95\% (1485 km). We initially set the size of symbols to be 16 bits and compare the average number of data packets needed to reconstruct the authentication payload under different PLR is given in Figure~\ref{figPLRvsNeededPkts}. The number of encoded packets increases with two factors: i) the size of the raw authentication payload. ii) the packet loss rate of the wireless channel. We also compare the minimal value of $r_0$ under different $PLR$ in Figure~\ref{figPLRvsNeededPkts}. The minimal value is defined as:
\begin{align}
    r^*_0 = \dfrac{N_{PLR}}{N_0}
\end{align}
where $N_{PLR}$ is the average number of non-repeated encoded packets needed to reconstruct the authentication payload properly. $N_0$ is the minimal number of encoded packets needed to reconstruct the payload where $PLR=0$. As in Figure~\ref{figPLRvsNeededPkts}, $r^*_0$ grows exponentially with the increase of PLR and does not seem to be affected by the size of raw payloads.
\begin{figure}
    \centering
    \includegraphics[width=0.8\linewidth]{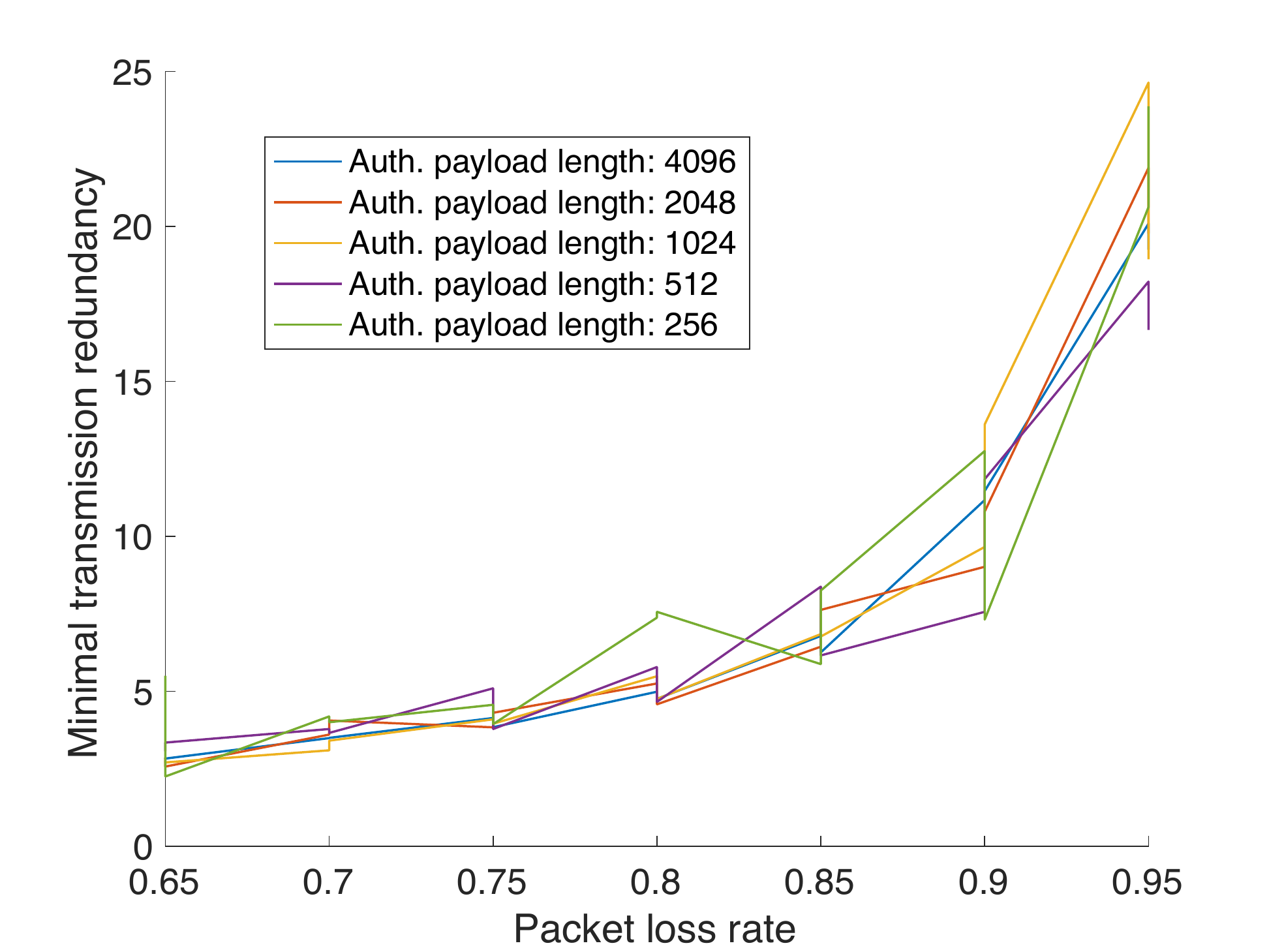}
    \caption{Transmission redundancy with different PLR and symbol size being 16 bits}
    \label{figPLRvsMinimalRedundancy}
\end{figure}

\begin{figure}
    \centering
    \includegraphics[width=0.8\linewidth]{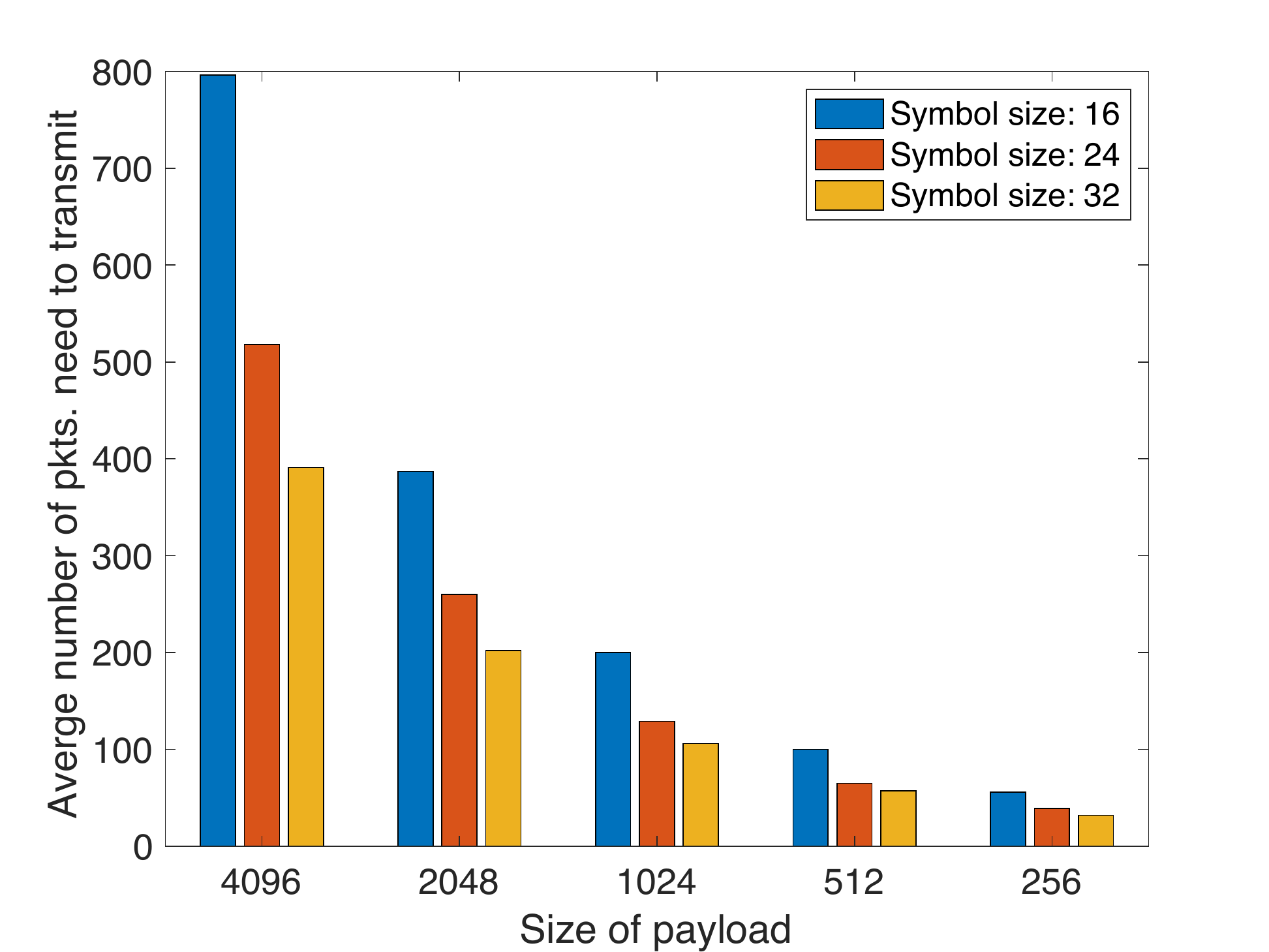}
    \caption{The average number of packets needed to properly deliver authentication payloads (the range of communication is less than 100 km).}
    \label{figAvgPktsNeededWithin100km}
\end{figure}

\begin{figure}[h]
    \centering
    \includegraphics[width=0.8\linewidth]{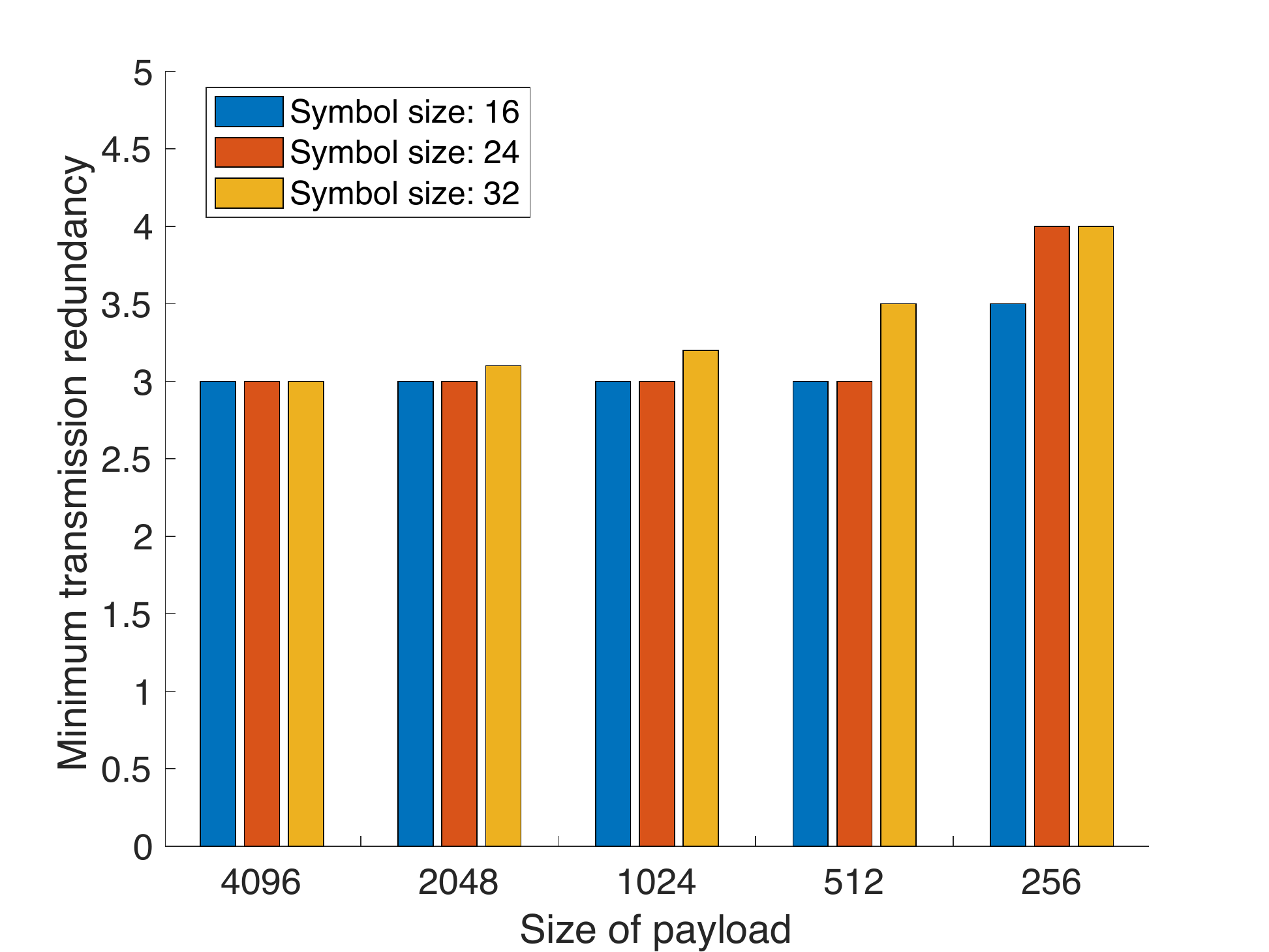}
    \caption{The average redundancy of data transmission to properly deliver authentication payloads (the range of communication is less than 100 km).}
    \label{figAvgRedundancyWithin100km}
\end{figure}

If we limit the range of UAS operation to 100 km, the maximum PLR becomes 68\%. Then we can simulate the performance of the authentication encoding scenario. We focus on the minimal redundancy and the average number of packets we need to transmit to deliver the authentication payload properly. As depicted in Figure~\ref{figAvgPktsNeededWithin100km} and \ref{figAvgRedundancyWithin100km}, large authentication payloads require more data packets to deliver properly, but the average transmission redundancy maintains relatively stable. In the meantime, we can reduce the packets needed to deliver the payload. The redundancy of transmission is relatively stable.

\subsection{Practical Configuration and Adaptation}

Suppose that we want to deploy this authentication framework into the real ADS-B transponders for UAS and regular aircraft. We can configure the size of the authentication payload to be 1024 bits and the symbol size to be 32 bits. In this way, authentication messages can be delivered and reconstructed within 120 ADS-B packets (less than one minute). As in Figure~\ref{figAvgRedundancyWithin100km}, the redundancy for transmission needs to be around 3.25. Therefore, we should provide $\ceil*{log_2 (1024\times3.25)/32} = 7 $ bits to store the encoding hints $L_h$. As we have $51~bits > 32 + 7 = 39(bits)$. Consequently, the de-facto ADS-B protocol is capable of carrying on the proposed authentication framework. Even if we use a 2048-bit authentication payload, we only need $\ceil*{log_2 (2048\times3.25)/32} = 8$ bits to store the decoding hints. We still have 43 bits to store the authentication symbols. Therefore, we believe that this authentication payload encoding algorithm is fully compatible with the de facto ADS-B protocol in air traffic control facilities.

\section{Conclusion}
\label{sectCC}
In this paper, we propose a secure framework for the safe integration of UAS into the national airspace. We first propose a blockchain enabled framework for the authentication of flight plans and exchanging authentication secrecy between UAS and ATM infrastructures in the preflight stage where reliable Internet connection is available to the UAS. We then provide a fountain code enhanced ADS-B protocol for the en-route authentication of UAS, and this protocol is seamlessly compatible with the de facto ADS-B ecosystem for ATC and ATM. More importantly, we theoretically prove the compatibility of this protocol in the de facto ADS-B systems.

\section*{Acknowledgment}

This research was partially supported by the Center for Advanced Transportation Mobility (CATM), USDOT under Grant No. 69A3551747125 and the National Science Foundation under Grant No. 1956193.

\bibliographystyle{IEEEtran}
\bibliography{GlobalComRef.bib}

\end{document}